\documentclass[preprint,a4,prd,aps]{revtex4}
\usepackage{amsmath}
\begin{document}                                                 
\newcommand{\be}{\begin{equation}}
\newcommand{\ee}{\end{equation}}
\newcommand{\ba}{\begin{eqnarray}}
\newcommand{\ea}{\end{eqnarray}}
\newcommand{\bc}{\begin{center}}
\newcommand{\ec}{\end{center}}
\newcommand{\vs}{\vspace*{3mm}}
\newcommand{\dis}{\displaystyle}
\newcommand{\bay}{\begin{array}}
\newcommand{\eay}{\end{array}}
\def\RN{Reis\-sner-Nord\-str\"{o}m }
\def\rc{\rho_{\rm crit}}
\def\rl{\rho_\Lambda}
\def\rt{\rho_{\rm tot}}
\def\ie{{\it i.e.\;}}
\def\lp{\ell_{\rm Pl}}
\def\mp{m_{\rm Pl}}
\def\tp{t_{\rm Pl}}
\def\tf{t_{\rm FP}}
\def\om{\Omega_{\phi}}
\def\oa{\Omega_{\Lambda}}
\def\ot{\Omega_{\rm tot}}
\def\tla{\widetilde{\lambda}_\ast}
\def\tom{\widetilde{\omega}_\ast}\
\def\luv{\lambda_\ast}
\def\guv{g_\ast}
\def\lir{\lambda_\ast^{\rm IR}}
\def\gir{g_\ast^{\rm IR}}
\def\lir{\lambda_\ast^{\rm IR}}
\def\gir{g_\ast^{\rm IR}}
\title{A class of renormalization group invariant 
scalar field cosmologies}
\author{Alfio Bonanno}
\email{Electronic address: abo@ct.astro.it}
\affiliation{INAF, Osservatorio Astrofisico di Catania, Via S.Sofia 78,
I-95123 Catania, Italy}
\altaffiliation[Also at ]{INFN, Sezione di Catania, via S.Sofia 73}
\author{Giampiero Esposito}
\email{giampiero.esposito@na.infn.it}
\affiliation{INFN, Sezione di Napoli, and Dipartimento di 
Scienze Fisiche\\
Complesso Universitario di Monte S. Angelo, Via Cintia,
Edificio N', 80126 Napoli, Italy}
\author{Claudio Rubano}
\email{Electronic address: claudio.rubano@na.infn.it}
\affiliation{Dipartimento di Scienze Fisiche and INFN,
Sezione di Napoli\\
Complesso Universitario di Monte S. Angelo,\\
Via Cintia, Edificio N', 80126 Napoli, Italy}
\altaffiliation[Also at ]{Istituto Nazionale 
di Fisica Nucleare, Sezione di Napoli}
\begin{abstract}
We present a class of scalar field cosmologies with a dynamically 
evolving Newton parameter $G$ and cosmological term $\Lambda$.
In particular, we discuss a class of solutions which are consistent 
with a renormalization group scaling
for $G$ and $\Lambda$ near a fixed point.
Moreover, we propose a modified action for gravity which includes the 
effective running of $G$ and $\Lambda$ near the fixed point. 
A proper understanding of the associated variational problem is 
obtained upon considering the four-dimensional gradient 
of the Newton parameter.
\end{abstract}

\maketitle

\section{introduction}
The recent discovery that Einstein gravity is most probably renormalizable
at a non-perturbative level \cite{ol1,ol2,ol3,frank}
has triggered many investigations on the 
possible consequences of these findings
in cosmology. In \cite{br1}, a cosmology of the Planck Era, 
valid immediately after the initial singularity, 
was discussed. In this model the Newton constant $G$ 
and the cosmological constant $\Lambda$ are dynamically  
coupled to the geometry by ``improving'' 
the Einstein equations with the renormalization group (hereafter RG) 
equations for Quantum Einstein Gravity \cite{mr}.
This modified Einstein theory is not affected by the 
horizon and flatness problems of the cosmological standard model. 

%Gravitational theories with variable $G$ have been discussed 
%in the context of the ``induced-gravity'' \cite{zee}
%and in the more general contex of the scalar the
In \cite{br2}, a similar framework has been extended 
%In \cite{br2}, the approch discussed in \cite{br1} 
%has been extended 
to the study of the large scale dynamics of the Universe. 
In this case a solution of the ``cosmic coincidence problem'' 
\cite{coscon2} arises naturally without 
the introduction of a {\rm quintessence} field, because
the vacuum energy density $\rho_\Lambda\equiv\Lambda/8\pi G$
is automatically adjusted so as to equal the matter energy density, 
{\it i.e.} $\oa=\Omega_{\rm matter} =1/2$ \cite{br2}. 
We shall call the models discussed 
in \cite{br1,br2} as {\it fixed point} 
(hereafter FP) cosmologies, or equivalently, 
RG-invariant cosmologies. 

In a nutshell, the {\it renormalization group improvement}  
consists in the modified Einstein equations
\be\label{ein}
R_{\mu\nu}-\frac{1}{2}g_{\mu\nu}R = -g_{\mu\nu}\Lambda(k)+
8\pi G(k) T_{\mu\nu}
\ee
where the Newton parameter $G$ and cosmological term $\Lambda$ 
are now dependent on the scale $k$, $k$ being 
the running cut-off of the renormalization group equation \cite{mr}.

%%%%%%%%%%%%%%%%%%%%%%%%%%%%%%
Gravitational theories with variable $G$ have been discussed 
in the context of the ``induced-gravity'' model \cite{zee}
where the Newton constant is generated by means of 
a non-vanishing vacuum expetation value of a scalar field.
However here the basic difference here is that the dynamical
content of the theory is not determined by a dynamical rearrangement
of the symmetry, but instead it is determined by the renormalization group
approach applied to the (quantum) Einstein-Hilbert lagrangian.
It is however interesting to notice the a dynamically evolving cosmological 
constant and gravitational interaction also appear
in very general scalar-tensor cosmologies \cite{cari97,carma}.
%%%%%%%%%%%%%%%%%%%%%%%%%%%%%%%%%%%%%%%%%%%%%%%%%%%%%%%%

This framework has been also applied in General Relativity in \cite{bh1},
in the dynamical context of a gravitational collapse,
and in \cite{bh2} for a Schwarzschild black hole.  
In cosmology, the dynamical evolution is instead determined by a set of 
renormalization group equations by means of the 
cut-off identification $k=k(t)$ which 
relates the energy scale of the running 
cutoff $k$ of the renormalization group, with the cosmic time $t$. 
In \cite{br2} it has been shown that, 
in a cosmological setting, the correct
cutoff identification is $k \propto t^{-1}$;
it is thus possible to determine $G(k(t))$ and $\Lambda(k(t))$ in
Eq. (\ref{ein}) once a RG trajectory is determined. 
The aim of this paper is to extend the results discussed in \cite{br1,br2}
to the case of a scalar field coupled to gravity.

Let us in fact assume that, besides the non-Gaussian fixed point discovered
in \cite{ol1} for pure gravity, the standard 
Gaussian fixed point is accessible in
perturbation theory in the scalar sector 
(this is actually the case for a free
scalar field as shown in \cite{perc1}, and it also emerges from the 
analysis of Ref. \cite{perc2} for a self-interacting scalar theory). Then,
a solution which is compatible with a possible RG trajectory for the scalar
sector must predict a simple renormalizable potential for spin-$0$ 
particles. We thus show that {\it there exists a class of solutions
for the familiar $\phi^{4}$ renormalizable potential}.
 
In addition, we also discuss a possible 
renormalization-group improvement at the level of the
Einstein-Hilbert Lagrangian itself. In this case, solutions for a
class of power-law self-interaction potentials are available only
for some specific values of the quartic self-interaction 
coupling constants.

The plan of this work is the following: in Sec. II we introduce 
the basic equations and present the scalar
field solution. In Sec. III we discuss the RG improvement
of the Einstein-Hilbert Lagrangian. 
Section IV is devoted to the conclusions.
\section{the model}
We now introduce the basic equations of the FP cosmologies for a scalar 
field matter component. 
Let us recall that the effective energy density and pressure of a generic 
scalar field read:
\ba\label{eq:dens}
&&\rho_\phi = {1\over 2}\dot{\phi}^2+V(\phi), \\[2mm]
&& p_\phi   = {1\over 2}\dot{\phi}^2-V(\phi),
\ea
respectively. In term of $\rho_\phi$ and $p_\phi$ 
the coupled system of RG improved evolution equations 
read
\begin{subequations}
\label{eq:system}
\ba
&&\left(\frac{\dot{a}}{a}\right)^2+\frac{K}{a^2}=\frac{1}{3}\Lambda+
\frac{8\pi}{3}G\rho_\phi, \label{1a}\\
&&\ddot{\phi}+3\frac{\dot{a}}{a}\dot\phi+V'(\phi)=0, \label{1b}\\
&&\dot{\Lambda}+8 \pi \dot G \rho_\phi=0, \label{1c}\\
&&G(t) \equiv G(k(t)), \;\; \Lambda(t) \equiv \Lambda(k(t)), \label{1d}
\ea
\end{subequations}
Eq.(\ref{1a}) is the improved Friedmann equation, 
Eq.(\ref{1b}) is the Klein-Gordon equation,
Eq.(\ref{1c}) follows from the Bianchi identities, 
and Eqs.(\ref{1d}) are determined from the 
renormalization group equations once the cutoff 
identification $k=k(t)$ is given.
We define the vacuum energy density $\rl$, the total energy density
$\rt$ and the critical energy density $\rc$ according to
\ba
&&\rho_{\Lambda}(t) \equiv \frac{\Lambda(t)}{8 \pi G(t)}, \\
&&\rho_{\rm{tot}}(t) \equiv \rho_\phi + \rl , \\
\label{eq:cdens}
&&\rho_{\rm{crit}}(t) \equiv \frac{3} {8 \pi G(t)} \left(\frac{\dot{a}}{a}
\right)^{2} ,
\ea
with $H \equiv \dot a / a$.
Hence we may rewrite the improved Friedmann equation (\ref{1a}) in the form
\be
\label{eq:IFE2}
\frac{\dot a^2+K}{a^2} = \frac{8 \pi}{3} G(t) \rt .
\ee
We refer the various energy densities to the critical density
(\ref{eq:cdens}):
\ba
&&\Omega_{\phi} \equiv \frac{\rho}{\rho_{\rm crit}}, 
\quad \Omega_{\Lambda} \equiv 
\frac{\rho_{\Lambda}}{\rho_{\rm crit}} , \\
&&\Omega_{\rm tot}= \Omega_{\phi} + 
\Omega_{\Lambda} \equiv \frac{\rho_{\rm tot}}{\rho_{\rm crit}} .
\ea
It follows from these definitions that
the Friedmann equation (\ref{eq:IFE2}) becomes
\be
\label{eq:kol}
K=\dot a^2 \; [\Omega_{\rm tot}-1].
\ee
For a spatially flat universe ($K=0$) we need $\rt=\rc$, as in
standard cosmology. In the following we shall discuss only the $K=0$ case. 
In order to solve the system (\ref{eq:system}) 
we consider the first three equations
in (\ref{eq:system}) without the RG equations (\ref{1d}). 
While in general (4) can be solved once $V(\phi)$ is given, we
shall see that the {\it perfect fluid} ansatz $p_{\phi}=w \rho_{\phi}$,
$w$ being a constant, is equivalent to assume a class of
power-law potentials $V(\phi) \propto \phi^{m}$.

We first consider the first three equations in (4) without the RG
equations (4d), and then we determine the solutions consistent 
with a given RG trajectory. The potential can be written as
\be\label{eq:pot}
V(\phi) = \frac{1}{2}\dot{\phi}^2 \Big ( \frac{1-w}{1+w} \Big ),
\ee
which shows that the value $w=-1$ should be ruled out, as
we will do from now on. By
substitution in the Klein--Gordon equation (\ref{1b}) we readily obtain
\be\label{eq:eu}
\rho_\phi = \frac{1}{1+w}\dot{\phi}^{2}
\equiv \frac{\cal M}{8\pi a^{3(1+w)}},
\ee
where ${\cal M}$ is an integration constant. 
By substituting into Eq. (\ref{1a}) we derive the following power-law 
solutions:
\begin{subequations}
\label{sol}
\ba\label{23a}
&&a(t) = \Big [\frac{3(1+w)^2}{2(n+2)}\;{\cal M}\; C\Big ]^{1/(3+3w)}
\;t^{(n+2)/(3+3w)}, \\[2mm]
&&\phi(t) = \Big(\frac{4(n+2)}{12\pi(1+w)Cn^2}\Big)^{1/2} 
t^{-n/2}, \label{23b}\\[2mm] 
&&G(t) = C\; t^{n} , \label{23c}\\[2mm]
&&\Lambda(t) = \frac{n(n+2)}{3(1+w)^2}\;\frac{1}{t^2}, \label{23d}
\ea
\end{subequations}
where $C$ is a constant 
and $n$ is a positive integer. For example, writing 
$a(t)=\alpha t^{\alpha}, \Lambda=\beta t^{-2}$ and expressing $G$ as
in (14c), Eq. (4a) yields, for $K=0$, a first-degree algebraic
equation for $\alpha$, which is solved by $\alpha={(n+2)\over 3(1+w)}$.
Equation (\ref{eq:eu}) is then integrated to get the result (14b).
As anticipated, the potential is also a power law, i.e.
\be\label{potential}
V(\phi) = \frac{1-w}{2+2w}
\Big ( \frac{12\pi(w+1)C}{n+2} \Big)^{\frac{2}{n}} 
(\frac{n}{2})^{\frac{2(n+2)}{n}} 
\;\;\;\phi^{\frac{2(n+2)}{n}}.
\ee
The RG equations (\ref{1d}) have not been used so far. What is the correct 
RG trajectory for a self-interacting scalar field coupled with gravity? 
Let us consider the RG-trajectory discussed in the introduction, where
in the deep UV region we must have the non-Gaussian 
fixed point \cite{ol1,ol2,ol3,frank} in the gravitational
sector, and  the Gaussian one in the scalar field sector.
In this case, {\it since the renormalized trajectory ends at} $(\luv,\guv)$,
the dimensionful quantities must run as
\be\label{rg}
G(k) = {\guv}/{k^2}, \;\;\;\; \Lambda(k) = \luv \;k^2
\ee
where $\guv$, $\luv$ are the dimensionless coupling $g(k)$ and $\lambda(k)$,
respectively, at the ultraviolet non-Gaussian 
fixed point $k\rightarrow \infty$.
The numerical values have been obtained in 
the analysis of \cite{perc1,perc2}
and read  $\guv \approx 0.31$, $\luv \approx 0.35$ approximately. 

The next step is to determine $k$ as a function of $t$. 
In \cite{br1} it was shown that the 
correct cutoff identification is given by 
\be\label{cuti}
k(t) = {\xi}/{t} .
\ee
Therefore, we see from (\ref{rg}) and from (\ref{cuti}) that 
$G= \guv \xi^{-2} t^2$ and $\Lambda =\luv  \xi^{2} t^{-2}$, therefore  
we must choose 
$n=2$ in (\ref{sol}) and $\xi^2 = 8/3(1+w)^2\luv$ in (\ref{cuti}). At last
the following {\it renormalization group invariant} 
(or fixed-point) solution is obtained:
\begin{subequations}
\label{sol2}
\ba\label{24a}
&&a(t) = \Big [\Big (\frac{3}{8}\Big )^2 (1+w)^4 \guv \luv \;{\cal M} 
\Big ]^{1/(3+3w)}\;t^{4/(3+3w)}, \\[2mm]
&&\phi(t) = \Big(\frac{8}{9\pi(1+w)^3\guv\luv}\Big)^{1/2} 
\; \frac{1}{t}, \label{24b}\\[2mm] 
&&G(t) = \frac{3}{8}(1+w)^2\guv\luv t^{2}, \label{24c}\\[2mm]
&&\Lambda(t) = \frac{8}{3(1+w)^2}\;\frac{1}{t^2}. \label{24d}
\ea
\end{subequations}
The solution, as far as $a(t)$, $G(t)$ and $\Lambda(t)$ are concerned, is 
basically the same as what already discussed in \cite{br1,br2} but 
in this case the potential reads
\be\label{pot2}
V(\phi) = \frac{9 \pi}{16}(1-w)(1+w)^2\guv\luv \; \phi^{4},
\ee
which is the standard renormalizable quartic self-interacting potential for 
a massless scalar theory. The role of $w$ is now clear: it allows a 
convenient parametrization of the solution in terms of the parameter $w$
instead of the self-interaction coupling constant in the potential.
It in fact measures the 
self-coupling strength $9(1-w)(1+w)^2\guv\luv/16$:
for $w=1$ ({\it stiff matter} equation of state) $V=0$ and 
$\phi$ is a free field, while for $0<w<1$, $\phi$ is an interacting field.  
For $w>1$ the theory is not bounded from below. 

Other properties of the solution (\ref{sol2}) 
have been extensively discussed
in \cite{br1,br2} and we shall not repeat this discussion here.
We simply point out that for the solution (\ref{sol2}), 
we have $\Omega_\phi=\Omega_\Lambda=1/2$ at any time.
\section{improving the action}
One of the striking properties of the renormalization group 
trajectory (\ref{rg}) is that the following 
relation holds:
\be\label{rela}
\Lambda = \frac{\guv\luv}{G} .
\ee
This fact has a deep meaning and is related with 
the possibility of reducing 
the number of coupling constants in a RG-invariant theory \cite{zimmer}.
What happens in our case is that near the
fixed point it is always possible to consider 
$\Lambda=\Lambda(G)$ and the effective
scaling is ruled only by $G$, for instance.
This fact suggests that a more fundamental approach 
should consider $\Lambda$ as a function of $G$ from the beginning, 
perhaps at the level of the action itself.

Let us therefore consider the action $S=S_{g}+S_{m}$, where $S_{m}$ is the
action for the matter field, and
\be\label{action}
S_{g} = \int_{M} d^4x\;\sqrt{-g}\;
\left(\frac{R}{G}-\frac{2\Lambda(G)}{G}\right),
\ee
where $M$ is the portion of space-time we have access to.
This is a well-defined {\it starting point} for promoting $G$ and
$\Lambda$ to the role of dynamical variables in a fully covariant
way. However, since the scalar curvature contains second derivatives
of the metric and $G$ is no longer constant, some extra care is necessary
to obtain a well-posed variational problem. Indeed, on denoting by
$\Gamma_{\; \mu \nu}^{\lambda}$ the Christoffel symbols, and 
defining \cite{fock}
\begin{equation}
w^{\alpha} \equiv g^{\mu \nu} \delta \Gamma_{\; \mu \nu}^{\alpha}
-g^{\alpha \nu}\delta \Gamma_{\; \mu \nu}^{\mu},
\label{(22)}
\end{equation}
variation of $S_{g}$ yields
\begin{eqnarray}
\delta S_{g} &=& \int_{M}{1\over G}\left({R\over 2}g^{\alpha \beta}
-R^{\alpha \beta}\right)\delta g_{\alpha \beta}\sqrt{-g}d^{4}x
+\int_{M}\left[-{R\over G^{2}}+{2\Lambda \over G^{2}}
-{2\over G}{d\Lambda \over dG}\right]
\delta G \sqrt{-g}d^{4}x \nonumber \\
&-& \int_{M}{\partial \over \partial x^{\alpha}}
\left({\sqrt{-g}w^{\alpha}\over G}\right)d^{4}x
-\int_{M}{G_{,\alpha}\over G^{2}}\sqrt{-g}w^{\alpha}d^{4}x.
\label{(23)}
\end{eqnarray}
Thus, even upon choosing variations $\delta g_{\mu \nu}$ and 
$\delta \Gamma_{\; \mu \nu}^{\lambda}$ such that $w^{\alpha}$
vanishes on the boundary of $M$ \cite{fock}, the variation of the action
functional $S_{g}$ does not reduce to the first line of Eq. (23),
because the fourth term on the right-hand side of Eq. (23)
survives. We are therefore {\it assuming} that the gravitational
part of the action is actually ${\widetilde S}_{g}$ such that
\begin{equation}
\delta {\widetilde S}_{g}=\delta S_{g}
+\int_{M}{G_{,\alpha}\over G^{2}}\sqrt{-g}w^{\alpha}d^{4}x.
\label{(24)}
\end{equation} 
The content of our postulate is non-trivial, since the two variations
do not differ by the integral of a total derivative, as is clear
from (22) and (24). As far as we know, such a crucial point had not
been previously appreciated in the literature. The explicit
construction of ${\widetilde S}_{g}$ itself is more easily obtained
upon using an Arnowitt--Deser--Misner space-time foliation. On using
the standard notation for induced metric $h_{ij}$, extrinsic
curvature $K_{ij}$, lapse $N$ and shift $N^{i}$ \cite{misner}
one can show that the action (here $K \equiv h^{ij}K_{ij}, \;
h \equiv {\rm det} \; h_{ij}$)
\begin{equation}
{\widetilde S}_{g} \equiv \int_{M}{(R-2\Lambda)\over G}
\sqrt{-g}d^{4}x
+2 \int_{M}(K \sqrt{h})_{,0}d^{4}x
-2 \int_{M}{f_{\; ,i}^{i}\over G}d^{4}x,
\label{(25)}
\end{equation}
where $f^{i} \equiv K \sqrt{h}N^{i}-\sqrt{h}h^{ij}N_{,j}$,
reduces to
\begin{equation}
{\widetilde S}_{g}=\int_{M}{N \sqrt{h}\over G}
\Bigr(K_{ij}K^{ij}-K^{2}+{ }^{(3)}R-2 \Lambda \Bigr)d^{4}x,
\label{(26)}
\end{equation}
where ${ }^{(3)}R$ is the scalar curvature of the spacelike
hypersurfaces which foliate the space-time manifold when the
${\bf R} \times \Sigma$ topology is assumed. The action (25) is
the $3+1$ realization of an action fulfilling the condition (24),
as can be seen upon using the Leibniz rule to re-express
$$
{1\over G}(K \sqrt{h})_{,0}={G_{,0}\over G^{2}}K \sqrt{h}
+\left({K \sqrt{h}\over G}\right)_{,0} , \;
{1\over G}f_{\; ,i}^{i}={G_{,i}\over G^{2}}f^{i}
+\left({f^{i}\over G}\right)_{,i}.
$$
On the other hand, the identity (26) shows that
${\widetilde S}_{g}$ is eventually cast in the desired form 
suitable for calculus of variations, which only involves the
induced metric and its first derivatives, as well as the 
undifferentiated Newton parameter.
 
At this stage, variation of ${\widetilde S}_{g}$ with respect to 
$g_{\mu \nu}$ leads to Eqs. (\ref{eq:system})
and variation with respect to $G$ gives an additional constraint equation 
(see also \cite{krori}):
\be\label{con}
-\frac{R}{G}+\frac{2\Lambda}{G}-2\frac{d\Lambda}{d G} =0.
\ee
This equation, jointly with Eq.({\ref{1c}) and the field equations yields
\be\label{con2}
2\Lambda = 8\pi G (\rho+3p) = 8\pi G\rho(1+3 w).
\ee
Such a formula is a new equation with respect to the analysis in 
Ref. \cite{br2}, and expresses a restriction {\it which only holds if the 
potential is renormalizable}.
By inserting the general solution (\ref{sol}) in (\ref{con2}) we have
\be\label{con3}
n = (1+3w).
\ee
In particular, for the case of interest $n=2$, and hence $w=1/3$, leading
in turn to the renormalizable potential 
\be\label{pot3}
V(\phi) =\frac{2 \pi}{3}\guv \luv \phi^{4}. 
\ee
The relevant property of this solution is that the effective 
strength of the interaction self-coupling is determined entirely by
the fixed point values $\guv$ and $\luv$. For a free scalar field
$\guv \luv \approx 0.11$ \cite{perc1} and this value does not change in a
significant way in the interacting case \cite{perc2}. Loop corrections are
then expected to be small and the leading tree-level form of the
potential (30) holds. We can thus regard the cosmology (18) with
$w=1/3$ as an exact solution of the modified Einstein action
${\widetilde S}_{g}+S_{m}$ which is consistent with a RG flow near
the non-Gaussian fixed point in the gravitational sector and the
Gaussian one in the matter sector.
\section{conclusion}
We have presented a class of power-law cosmologies 
with variable $G$ and $\Lambda$ in the case
of a scalar field matter component, Eq.(\ref{sol})
(cf. important previous work in Ref. \cite{sasha} on scalar fields coupled
to gravity within the framework of renormalization group 
equations). We have then extended the FP cosmology 
presented in \cite{br1} by including the 
RG evolution Eq. (\ref{rg}) in the general solution (\ref{sol}). 
Last, we have presented a new RG-improvement at the level of the action
which picks out a specific self-interaction strength 
value for the scalar field potential.   
The scalar solution (\ref{sol2}) with $w=1/3$ can, at best, be considered 
only a toy model of the initial state of Universe. 
However, it may be helpful in understanding a more 
complete framework where the dynamical evolution of the gravitational field 
and the matter field near the initial singularity 
is consistent with RG scaling law of the renormalized 
theory near a fixed point.    
\section*{Acknowledgements}
The authors are indebted to the INFN and Dipartimento di Scienze Fisiche
of Naples University for financial support.
We thank M. Reuter for important suggestions on this work. 
A. B. acknowledges the warm 
hospitality of the University of Naples where part of this work was written.
The work of G. Esposito has been partially supported by PRIN
2002 {\it SINTESI}.

\end{document}